# EFFECTS OF AEROSOL AND MULTIPLE SCATTERING ON THE POLARIZATION OF THE TWILIGHT SKY

O.S. Ugolnikov[1], O.V. Postylyakov[2], I.A. Maslov[1]

[1]Space Research Institute, Profsoyuznaya st., 84/32, Moscow, 117997, Russia
[2]Institute of Atmospheric Physics, Pyzhevsky per, 3, Moscow, 109017, Russia
E-mail: ugol@tanatos.asc.rssi.ru

**Abstract.** The paper contains the review of a number of wide-angle polarization CCD-measurements of the twilight sky in V and R color bands with effective wavelengths equal to 550 and 700 nm. The basic factors effecting (usually decreasing) on the polarization of the twilight sky are the atmospheric aerosol scattering and multiple scattering. The method of multiple scattering separation is being considered. The results are compared with the data of numerical simulation of radiation transfer in the atmosphere for different aerosol models. The whole twilight period is divided on the different stages with different mechanisms forming the twilight sky polarization properties.

## 1. Introduction

The efficiency of the twilight observations for the atmosphere investigations was paid much attention long time ago and such early as in 1923 the twilight probing method became the subject of detailed analysis [1]. The difference with the daytime period, when the sky background is being formed by the light scattering in the lower troposphere layers, is accelerating elevation of the Earth's shadow over the observer and, thus, fast increasing of the effective scattering altitude. This gives the possibility of the retrieval of the scattering and absorption coefficient vertical profile, which can be useful for the investigations of aerosol and small admixtures in the atmosphere.

However, in the same paper [1] the basic problem of this method was specified. This problem is related with the solar light experienced two or more acts of scattering in the atmosphere, or multiple scattering. The process takes place in the dense lower layers of the atmosphere, and this twilight background component can be even brighter than the light scattered one time in the upper atmosphere. The reduction of multiple scattering based on some observational or theoretical methods is quite difficult task and it became the reason of sufficient differences of results obtained by different authors [2]. The possibility of exact numerical analysis appeared only in the last years.

In the paper [3] the high possible efficiency of polarization measurements of the twilight sky was noticed. The polarization data help to separate the single and multiple scattering, since the polarization of multiply scattered light is lower. This became the base of the method of single and multiple scattering separation suggested in [4]. Polarization measurements are effective for atmospheric aerosol sounding, and now POLDER space instrument [5] and AERONET ground network had started such observations. But since the aerosol scattering has the depolarization properties too, it can be mixed up with multiple scattering, this situation unfortunately taking place in a number of papers in 20$^{th}$ century.

In this paper we will consider the results of polarization measurements of the twilight sky in different spectral regions, compare it with the data of numerical simulation and establish the relation between the sky polarization evolution with the Sun depression and the contribution of atmospheric aerosol and multiple scattering.

## 2. Observations

The polarization twilight sky observations were conducted in 2002-2003 at South Laboratory Of Moscow Sternberg Astronomical Institute, situated in the Crimea, Ukraine. The observation device was the CCD-camera with short-focus lens and rotating polarization filter. The device axis was directed to the zenith, and the field of view was about 8.4x6.3 degrees. It was possible to hold the measurements from the day, when the Sun was above the horizon, till the nightfall. The procedure of star identification with the Tycho-2 star catalogue [6] made possible to exclude the stars up to 12$^m$ and to hold the photometry of bright stars for atmospheric transparency estimation at nighttime images. During the sessions in November-December 2002 and April 2003 the observations were carried out in V wide color band with the effective wavelength equal to 550 nm, and in April-June and September 2003 the twilight background was measured in R band with effective wavelength equal to 700 nm. The observations were similar to UBVR-measurements conducted in the same place in summer 2000 and described in [4] but with the more field of view (about 29x22 degrees).

## 3. Radiation transfer calculations and models

A simulation of the twilight polarized radiance was performed by using a Monte-Carlo linear radiation transfer model MCC++ [7]. This model calculates the radiation transfer with account of polarization in the spherical shell atmosphere taking into account all orders of scattering, aerosol and Rayleigh scattering, aerosol and gaseous absorption and surface albedo. The code was validated against other radiation transfer model for the twilight observations [8]. Calculations were carried out for the solar depressions under horizon up to 6° for the same wavelengths where 2000-2003 observations were made: 360, 440, 550 and 700 nm.

(a)

| Kind of particle | Dust | Water-soluble | Oceanic | Soot |
|---|---|---|---|---|
| Function of size distribution | lognormal | lognormal | lognormal | lognormal |
| Parameters | 0.5 mkm, 2.99 | 0.005 mkm, 2.99 | 0.3 mkm, 2.51 | 0.0118 mkm, 2.00 |

(b)

| Kind of aerosol | Dust | Water-soluble | Oceanic | Soot |
|---|---|---|---|---|
| Urban | 17% | 61% | 0% | 22% |
| Continental | 70% | 29% | 0% | 1% |
| Maritime | 0% | 5% | 95% | 0% |

**Table 1.** Kinds of aerosol particles (a) and percentage of aerosol particles in different kinds of aerosol (b) defined in [9].

The calculations were made for gaseous atmosphere with Rayleigh scattering matrix and for three standard aerosol models with microphysical characteristics proposed by WMO [9] and given in the Table 1: the urban, continental and maritime types of aerosol from 0 to 10 km and stratospheric aerosol above 10 km for all three cases. The phase matrix, absorption and scattering cross sections of aerosol were calculated by using Mie theory. The MODTRAN aerosol extinction profile at 600 nm, corresponding to surface visible range 50 km and background stratospheric aerosol, was used. Rayleigh extinction profile at 550 nm is taken from [10], the wavelength dependence was assumed as $\lambda^{-4}$, and the depolarization factor was ignored. The Krueger model of atmospheric ozone with total content 345 DU was used. Lambertian surface albedo was assumed to be equal to 0.5.

## 4. Multiple scattering separation method

A method of single and multiple scattered light separation using observations of polarization of twilight sky is described in details in [4]. The simplified idea of the method follows. When the Sun is situated near the horizon, the maximum polarization point of single scattering should be situated about 90° from it (or exactly 90° if the scattering is molecular) or near the zenith. And as the Sun is depressing, this point follows it by the vertical to the glow segment, remaining at the constant angular distance from the Sun (since the effective scattering altitude is not yet changing rapidly at the sunset moment). The maximum polarization point for multiple scattering at the sunset is expected to be near the zenith too, but as the secondary light source is the whole sky and basically, the glow area, and not directly related with the Sun, this point should be practically immovable remaining near the zenith. And thus, the position of the maximum polarization point for the total sky background will be the reflection of the changes of single and multiple scattering balance.

Figure 1 shows the dependence of the maximum polarization point zenith distance, $z_P$, on the sun depression under horizon, *h* (this value is negative at the daytime) for one date of simultaneous observations in 550 and 700 nm in 2000 [4]. The dashed line is corresponding to the case of pure Rayleigh single scattering. We can see that at the daytime and light phase of twilight until the Sun depression about 5°, the maximum polarization point really follows the Sun, but the velocity of such motion is lower a bit than for the case of pure single scattering. It points to almost constant ratio of single and multiple scattering at this time. And then this point is turning back to the zenith, and polarization profile of the solar vertical becomes symmetric relatively the zenith. It can be related only with the single scattering light disappearing on the background of multiple scattering.

A mathematical procedure of the multiple scattering separation method is described in [4]. This procedure uses the polarization data of light incoming from a few near-zenith directions for a few solar depressions during twilight, calculating the derivative values of sky polarization on the solar depression and zenith distance of the observation point. Method works if the contribution of single aerosol scattering is small (the influence of aerosol on multiple scattering can be sufficient and it is not a problem). As it was shown in [4] and will be visible below, it is true for the most part of cases at 550 nm and shorter wavelengths, where the effective single scattering at light twilight period takes place in the stratosphere. But at 700 nm the single troposphere aerosol scattering is sufficient and the exactness of the method decreases.

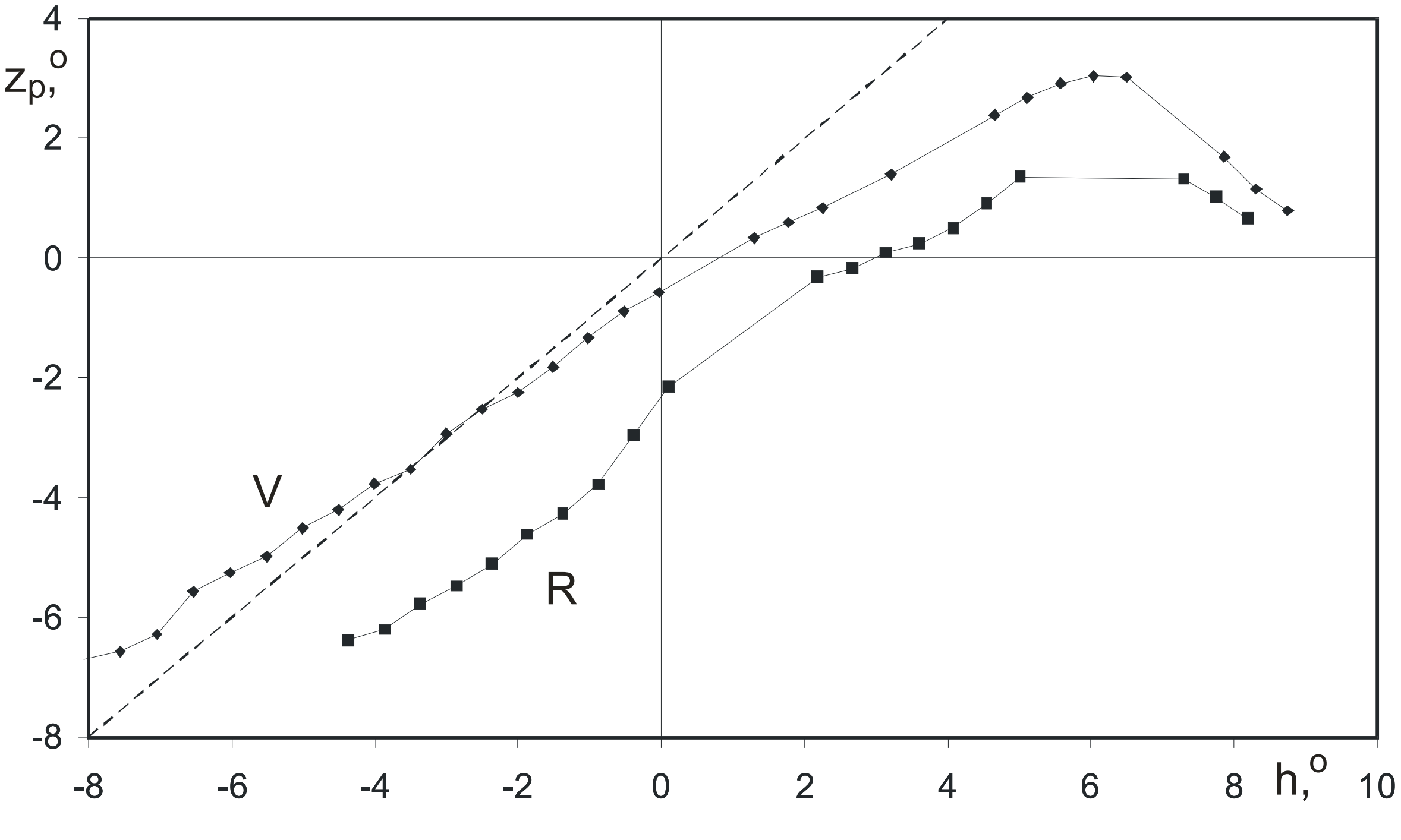

**Figure 1.** The motion of the maximum polarization point with the Sun depression in the evening of 06.08.2000 in V and R color bands (the dashed line corresponds to the case of pure Rayleigh single scattering).

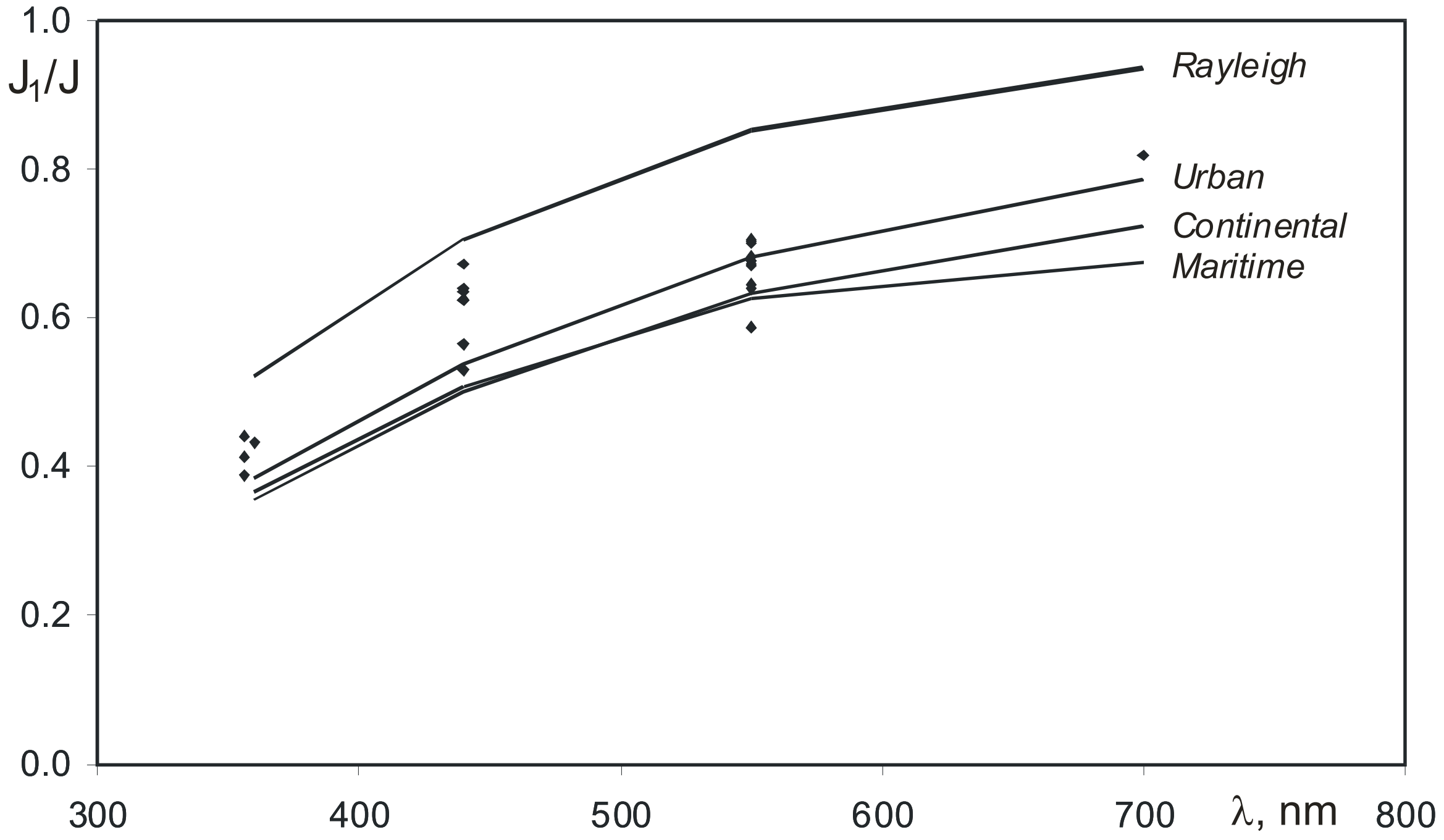

**Figure 2.** The contribution of single scattering in the twilight sky background at the zenith at the moment of sunrise (sunset) obtained by the observational method (dots) and radiation transfer simulation (lines).

The application of this method to the results of observations in summer 2000 [4] had shown that single scattering contribution decreases for the shorter wavelengths. Figure 2 shows the values of single scattering contribution at the zenith at the moment of sunrise or sunset (dots) compared with the results of calculations by the MCC++ code for different aerosol models (lines). Figure shows the good agreement of results of the observational method and radiation transfer simulations.

## 5. Observation results

### 5.1. Results in V band (550 nm)

Figure 3 shows the dependencies of sky polarization at the zenith, $p$, on the solar depression under the horizon, $h$, for a sample of 6 observation dates in 2000-2003 compared with the ones numerically simulated by the MCC++ model (bold lines). For any observation date, the twilight period can be divided to three sub-periods with different behavior of sky polarization. From the sunset until the sun depression about 5° the polarization is almost constant, then it begins to decrease and after the Sun depression about 9° in the moonless sky the polarization becomes constant again (it takes place until the depressions about 12-13° when the night sky contribution becomes noticeable). The nature of this evolution is related with the changes of the ratio of single and multiple scattering contribution, which is almost constant at light period of twilight, then the contribution of single scattering is decreasing completely vanishing at the Sun depression about 9°. In the same time the values of polarization of the single and multiple scattering components themselves are changing very slowly. Here we should note that by the same reason the twilight sky color has the similar behavior, getting blue at the time of polarization decrease [4].

The most interesting fact is that all curves in the figure are quite similar and just shifted one from another, in spite of the fact that observations were made in different years and seasons. It is true not only for dates shown in the figure, but for all ones in V band too. The same may be said about the theoretical curves. The value of sky polarization at the zenith at the light twilight period is unequivocally related with the one at the dark period when the twilight background consists of multiple scattering. It points to the fact that in V color band the changes of the atmospheric aerosol influence on the twilight sky polarization only through the multiple scattering component, and its influence being the same for all twilight stages. This fact is confirming the assumption about low contribution of single aerosol scattering in yellow and shorter-wave spectral regions [4]. It is also pointed by the constant value of sky polarization during the whole light twilight period with wide range of effective scattering altitudes.

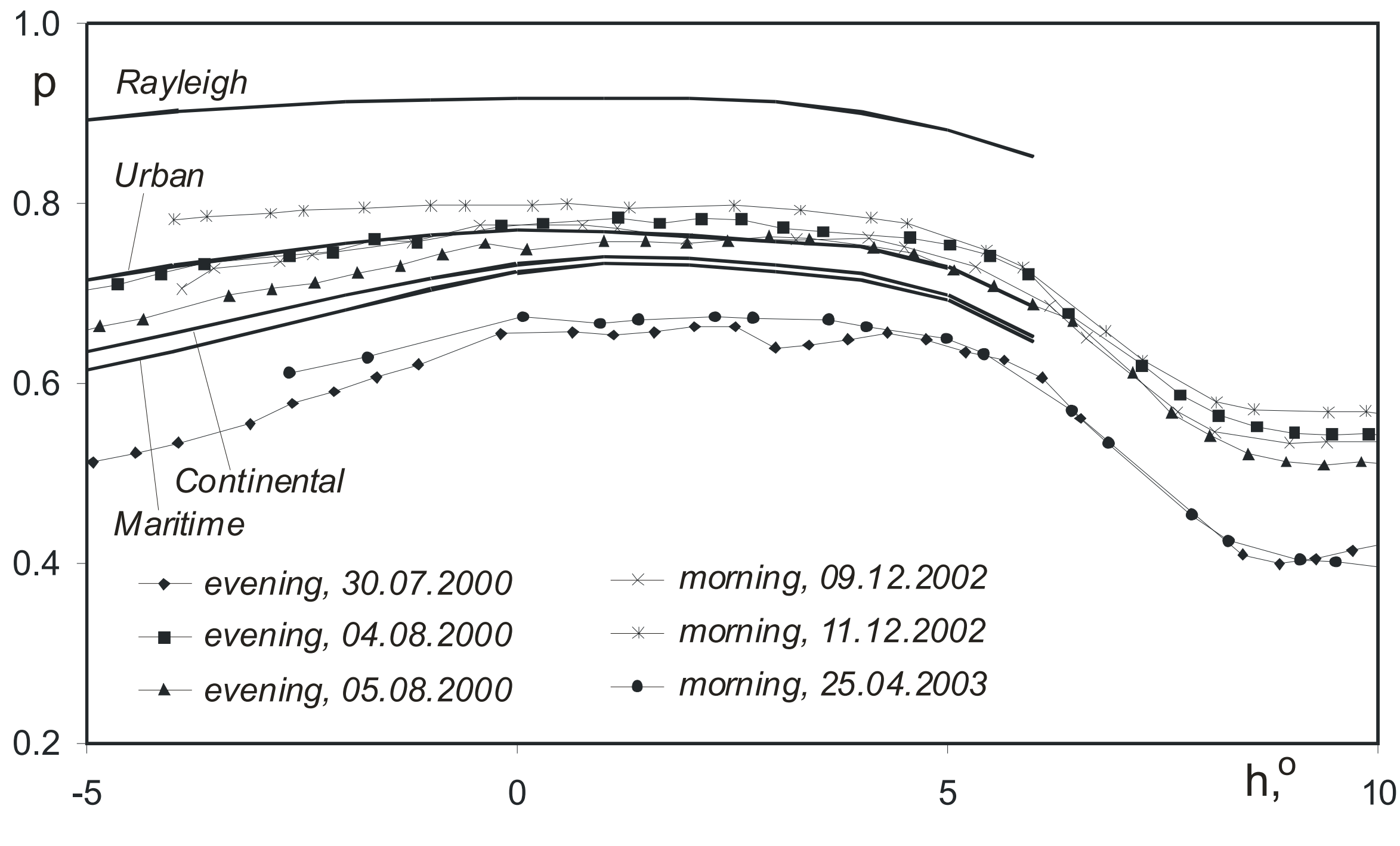

**Figure 3.** The dependencies of the sky polarization at the zenith on the Sun depression under horizon for different observation dates in V band (550 nm) compared with the results of numerical simulation of radiation transfer (bold lines).

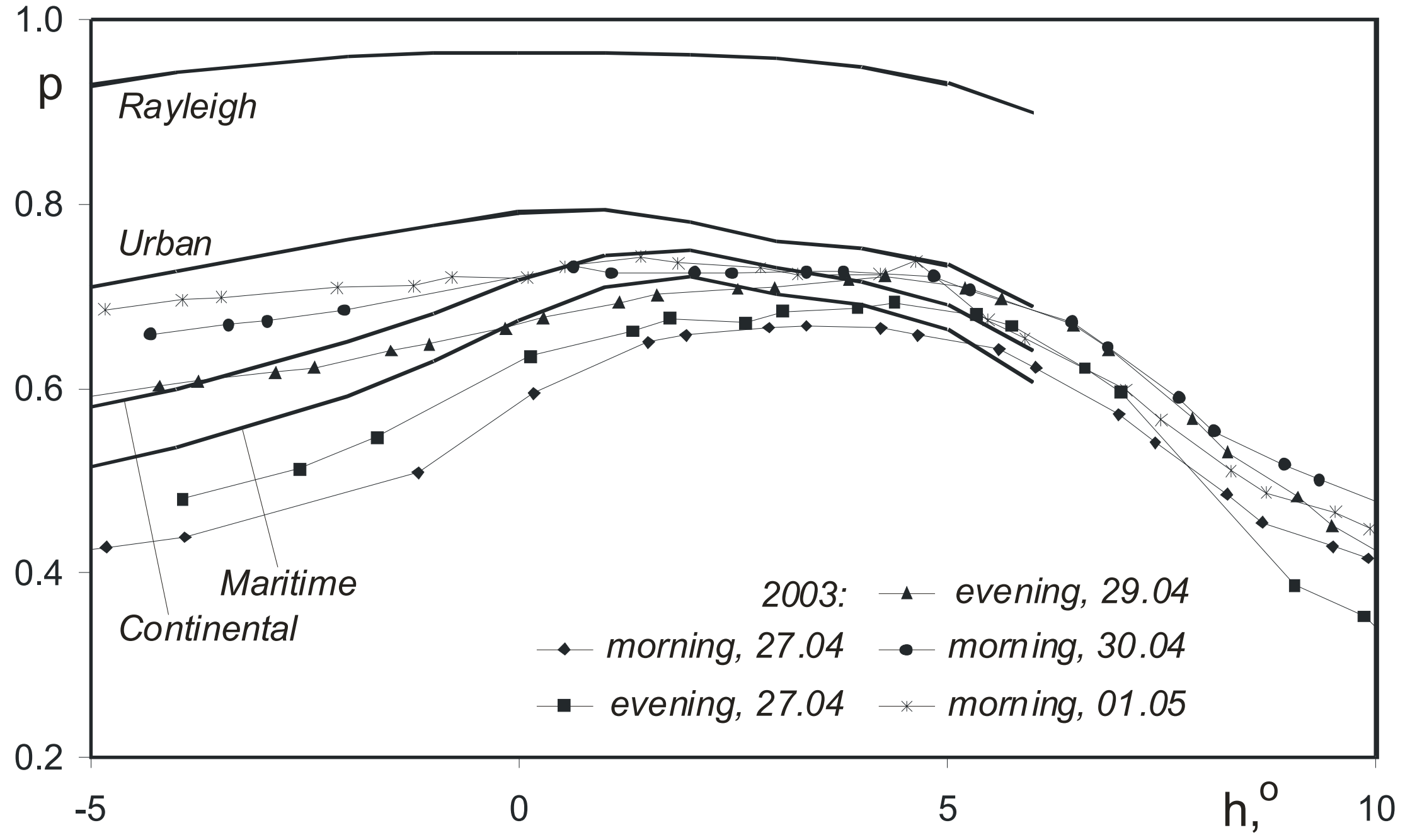

**Figure 4.** The dependencies of the sky polarization at the zenith on the Sun depression under horizon for different observation dates in R band (700 nm) compared with the results of numerical simulation of radiation transfer (bold lines).

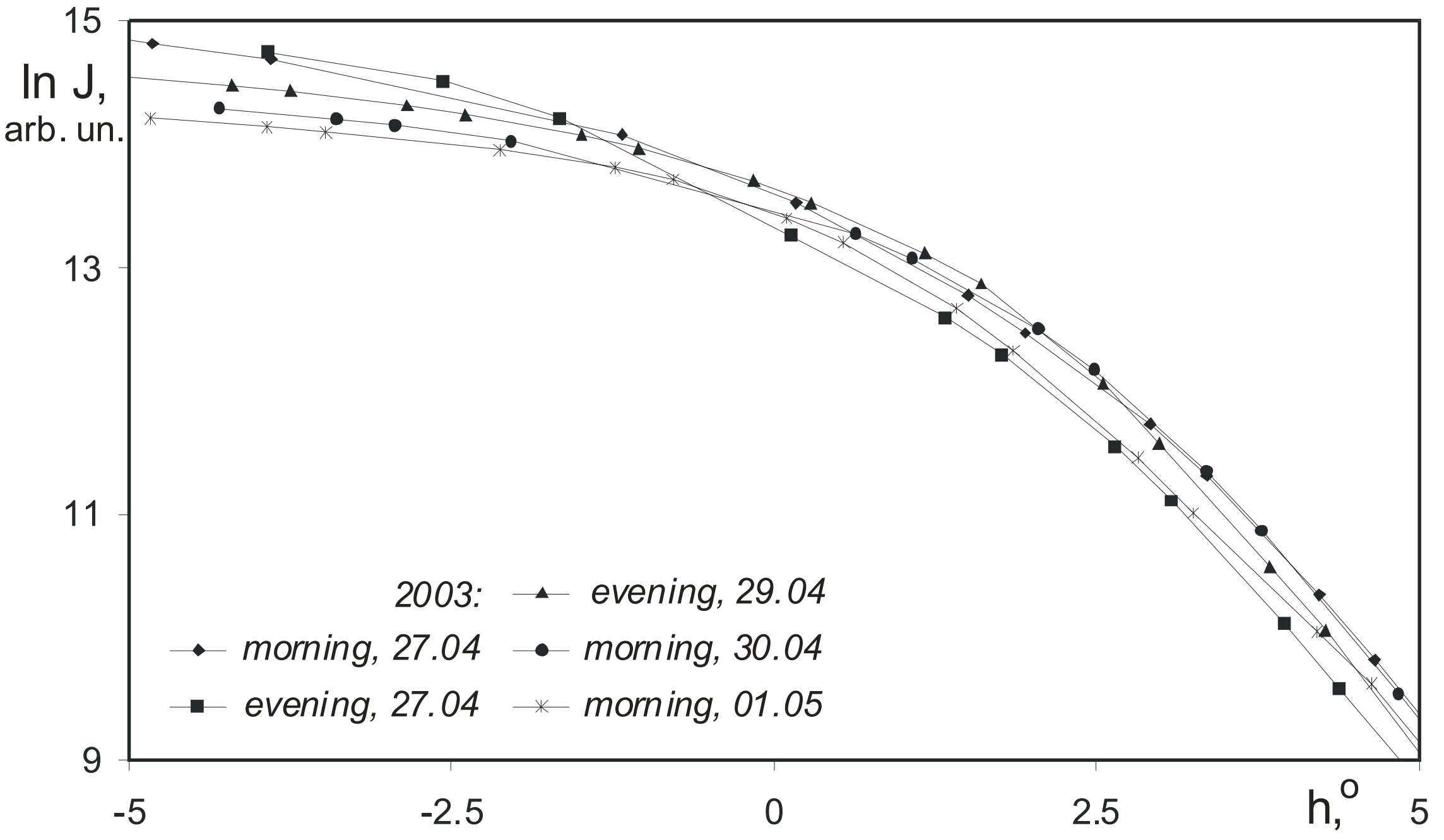

**Figure 5.** The dependencies of the sky brightness at the zenith on the Sun depression under horizon for different observation dates in R band.

The traces of single aerosol scattering appear only at $h<0$, or when the Sun rises above the horizon, for the dates with weak sky polarization and low atmosphere transparency. This time the aerosol in the lower layers of the atmosphere is becoming radiated by straight solar light without strong absorption. It is visible in Figure 3 by the polarization decrease at $h<0$ for these dates. The effect is not so strong, but as we can see below, in R color band, where the atmospheric absorption and effective scattering altitude are lower, it will be sufficient even during the twilight period at $h>0$.

The observational data for the dates with pure atmosphere and high polarization are in good agreement with the dependencies obtained by radiation transfer simulations, especially for urban aerosol model. But we should remember that this model data are very sensitive to the parameters and concentration of tropospheric aerosol forming the multiple scattering component and its small changes will move the curves in Figure 3 up or down practically without changing their shapes.

5.2. Results in R band (700 nm)

Figure 4 shows the same dependencies of sky polarization at the zenith on the depression of Sun for the sample of 5 observation dates at 700 nm. Despite of the fact that these observations were made during the 5 days period in 2003, the sufficient differences between the results are easy to see. The curves of the mornings of 30.04 and 01.05 are similar to the ones observed in V band and shown in Figure 3, but at other dates the polarization decreasing takes place not only at the daytime, but in the light twilight period until the depression of Sun about 3° too. Just after that the twilight sky polarization becomes principally the same for the dates shown in the figure. After the Sun depression about 5° the polarization fall related with the increase of multiple scattering contribution starts, but there is no plateau at the depressions about 10° like the one in V band. In the moonless conditions the polarization is monotonously decreasing until the nightfall. It is related with less intensity of multiple scattering and more intensity of night sky background in this spectral region.

Figure 5 shows the dependencies of sky brightness at the zenith on the depression of Sun for the same sample of observations in April – May 2003. We can see that in the daytime period ($h<0$) the brightness excess appears for the same dates when polarization is decreasing. This is obviously related with single scattering at the aerosol particles in the troposphere, the concentration of which is variable from day to day. But the effect of brightness excess appears only at $h<0$, although the polarization decrease at these dates is visible also during the twilight at $h<3°$. To explain it, we should remember that the aerosol reveals itself not only by additional scattering, but also by additional absorption. In the light twilight period these effects compensate each other, twilight sky brightness is almost independent on the aerosol contribution, which can be detected only by polarization measurements. When the Sun is above the horizon, the influence of aerosol absorption is less and the brightness excess appears.

In the Figure 6 we can see the solar depression dependency of correlation coefficient $C_{JP}$ between brightness logarithm and sky polarization calculated basing on the whole observational data in 2002-2003 in V and R bands. Being absent at the moment $h$=0, the anti-correlation of brightness logarithm and polarization in R band rapidly appears with sunrise, reaching 0.95 when the Sun is 3° above the horizon! In V band, where the absorption is stronger, the anti-correlation is less and it appears only at the Sun elevation about 3°.

Strong and variable influence of absorption does not give the possibility of direct measurement of the contribution and polarization of single aerosol scattering being based on this correlation and assuming that its intensity equal to the observed brightness excess.

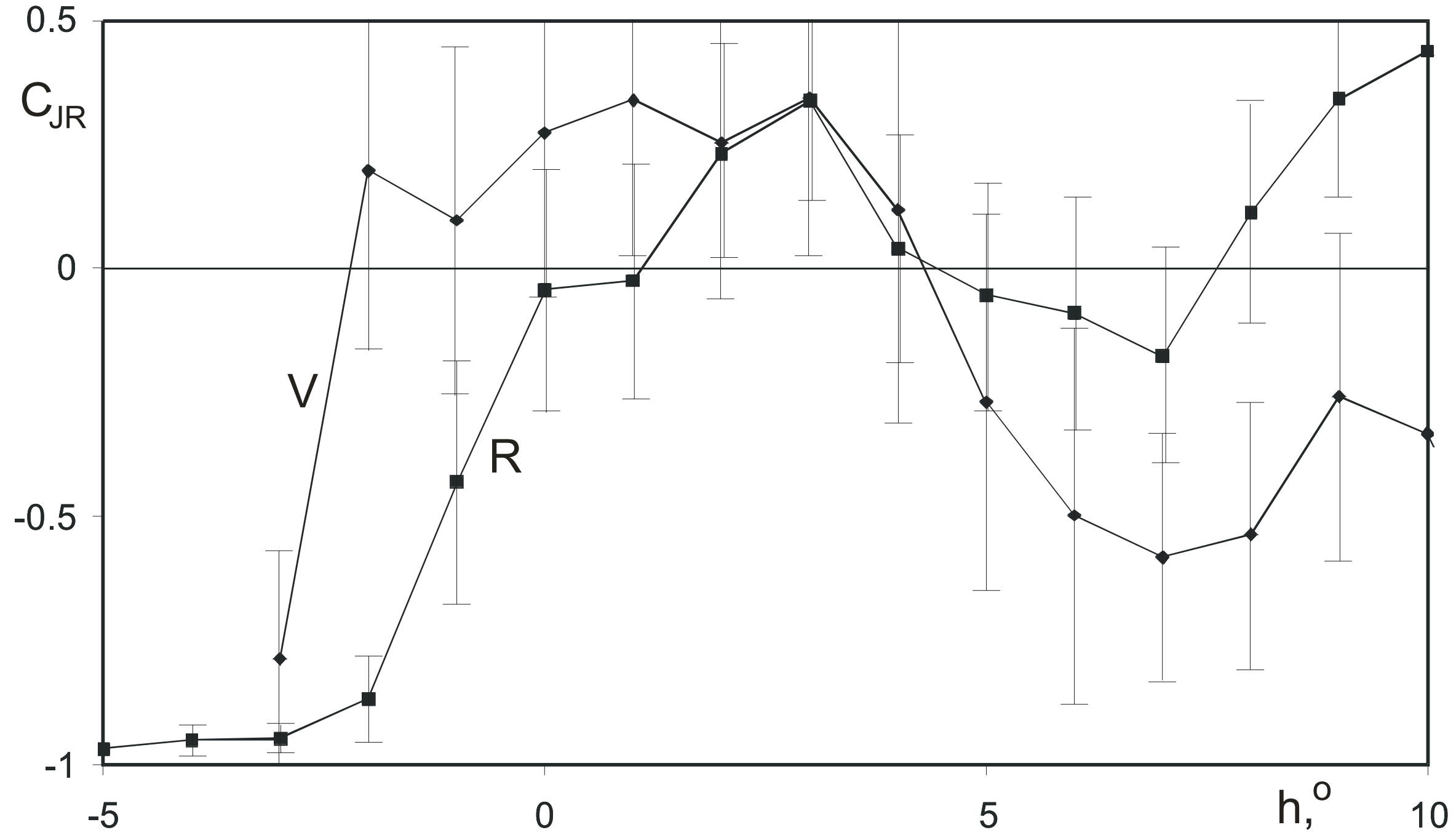

**Figure 6.** The dependency of correlation coefficient between sky brightness logarithm and sky polarization on the Sun depression under horizon at the zenith for 2002-2003 observations in V and R color bands.

Bold lines in the Figure 4, as in the Figure 3 for V band, are corresponding to the results of numerical simulations for gaseous atmosphere and three standard aerosol models. The difference between shapes of theoretical and observational curves as well as the difference of observational curves with each other is worthy of note. As against the multiple scattering, which is just shifting the curves up or down, the single aerosol scattering that is sufficient in R band changes also their shape that thus becomes a function of the vertical aerosol distribution.

## 6. Conclusions

The primary goal of this work is a qualitative analysis of the observed polarization features of the twilight sky for different wavelengths and twilight stages from the day till night. It was made to find out which depolarizing factor – multiple scattering or atmospheric aerosol – is the principal in this or that situation. This question is important for choose of the observational technique and the method of aerosol properties retrieval based on the twilight sky polarization data.

For any wavelength in visible part of spectrum and for any amount of Sun depression under horizon the atmospheric aerosol makes sufficient contribution to the intensity of multiple scattering component. But this is difficult to use in the retrieval methods. The observable signs of single aerosol scattering appear at definite Sun depression (or elevation) and can be observed only when the Sun is higher than this level. And just in red (and infrared) part of spectrum this period covers the light stage of twilight. The dependency of polarization on the Sun depression becomes the function of vertical aerosol distribution, and possibility of reverse task solution appears. At the same wavelengths the multiple scattering contribution decreases causing less problems for retrieval algorithms. Shorter wavelengths data are simpler for qualitative and quantitative analysis [4], but the contribution of single aerosol scattering at those wavelengths is many times weaker.